\documentstyle[12pt]{article}
\begin{document}

\title
{
Local-field effect in atom optics of two-component Bose-Einstein condensates
}
 
\author
{K.V.Krutitsky
\thanks{
		 Permanent address:
		 Ulyanovsk Branch of Moscow Institute of Radio
		 Engineering
		 and Electronics of Russian Academy of Sciences,
		 P.B.9868,
		 48, Goncharov Str., Ulyanovsk 432011, Russia;
		 e-mail: kostya@spock.physik.uni-konstanz.de, ufire@mv.ru
},
K.-P.Marzlin
\thanks{e-mail: Peter.Marzlin@uni-konstanz.de}
and J.Audretsch
\thanks{e-mail: Juergen.Audretsch@uni-konstanz.de}
\\
Fakult\"at f\"ur Physik,
Universit\"at Konstanz, Fach M 674,\\
D-78457 Konstanz, Germany
}

\date{}

\maketitle

\begin{abstract}
Starting from the first principles of nonrelativistic QED we have developed 
the quantum theory 
of the interaction of a two-component ultracold atomic ensemble with the 
electromagnetic field of vacuum and laser photons. The main attention 
has been paid to the consistent consideration of dynamical dipole-dipole 
interactions in the radiation field. Taking into account
local-field effects we have derived the system of Maxwell-Bloch equations.
Optical properties of the two-component Bose gas are investigated.
It is shown that the refractive index of the gas is given by
the Maxwell-Garnett formula.
All equations which are used up to now for the description
of the behavior of an ultracold atomic ensemble in a radiation field
can be obtained from our general
system of equations in the low-density limit. Raman-Nath
diffraction of the two-component atomic
beam is investigated on the basis of our general system of equations.
\end{abstract}

\newpage

%---------------------------------------------------------------------------
\section{Introduction}
%---------------------------------------------------------------------------

In recent years a great attention has been paid to the investigation of
two-component Bose-Einstein condensates (BEC). A two-component BEC can consist
of spatially separated identical atoms, or it can be a binary mixture of 
different alkali atoms, for instance, $^{87}$Rb--$^{23}$Na, or different isotops
like $^{87}$Rb--$^{85}$Rb, or different hyperfine states of the same alkali
atoms. A number of phenomena in two-component BECs, which are not possible
in single-component BECs, has been theoretically predicted and some
of them have been observed in experiments. It has been shown that the BEC
in a double-well potential can oscillate between the wells by quantum 
coherent atomic tunneling~\cite{JGT,MS}. Oscillations of this kind
can take place also in a two-component BEC, which consists of the same atoms
in different internal states~\cite{it}. Due to the nonlinearity arising
from atom-atom interactions, the oscillations are expected to be supressed
when the population difference of components exceeds a critical value in a process
known as macroscopic quantum self-trapping (MQST)~\cite{MS}.
However, in the process of collisions between the condensate and noncondensate
atoms MQST decays away~\cite{RW}. The dynamics of spatial separation of
two-component BEC has been studied in 
papers~\cite{EPZ,PHTS}.

In the present paper we shall investigate optical properties of two-component
BECs interacting with off-resonant laser radiation and develop mathematical
formalism for nonlinear atom optics with two-component condensates. Nonlinear
atom optics with single-component condensates is a rather well studied
subject. In papers~\cite{LLC,ZHA94a,WAL97,KBA99,KBA00} different
mathematical formalisms for the description of nonlinear phenomena in atom
optics of single-condensates has been proposed. Optical properties of
the single-condensates subject to the influence of off-resonant laser
radiation have been investigated 
in papers~\cite{WAL97,KBA99,KBA00,MRRMM}.
However, to our knowledge, nothing has been yet done in this direction for
multicomponent condensates. Following the ideas, presented in our previous 
papers~\cite{KBA99,KBA00}, we shall derive the system of Maxwell-Bloch
equations for nonlinear atom optics of two-component BECs. As an application
of our general theory we shall consider a diffraction of two-component
atomic beam from a standing light wave and discuss the specific
features of this phenomenon, which does not take place in the analogous
single-component process.

%--------------------------------------------------------------------
\section
{
The Hamiltonian for the two-component condensate interacting with photons
}
%--------------------------------------------------------------------

We consider a system of ultracold atoms which is a mixture of two species 
of two-level atoms with masses $m_1$, $m_2$, transition frequencies 
$\omega_1$, $\omega_2$, and matrix elements of the transition dipoles 
moments $d_1$, $d_2$. We shall describe such a system in terms of matter field
operators. Let 
$|g_j\rangle$ and $|e_j\rangle$, $j=1,2$ are the vectors of the ground and
excited states of the quantized atomic fields. Then the corresponding
annihilation operators of the atoms in these internal states are
$\hat \psi_{gj}$ and $\hat \psi_{ej}$.
Matter field operators are assumed to satisfy to the bosonic equal time
commutation relations and the operators of different components are assumed
to commute.

The Hamiltonian of the second quantized atomic field interacting with the
photons in the multipolar formulation of QED and in the electric dipole
approximation can be written down in the following manner
\begin{eqnarray}
\label{ham-second-quant}
\hat H
&=&
\hat H_A + \hat H_F + \hat H_{AI} + \hat H_{AF}
\;,
\\
\hat H_A
&=&
\sum_{j=1}^{2}
\left[
     \sum_{s=g,e}
     \int d{\bf r}\,
     \hat{\psi}_{sj}^\dagger({\bf r},t)
     \left(
          - \frac{\hbar^2 \nabla^2}{2m_j}
     \right)
     \hat{\psi}_{sj}({\bf r},t)
     +
     \int d{\bf r} \,
     \hat{\psi}_{ej}^\dagger({\bf r},t)
     \hbar \omega_j \hat{\psi}_{ej}({\bf r},t)
\right]
\;,
\nonumber\\
\hat H_F
&=&
\sum_{{\bf k}\lambda}
\hbar \omega_k
\hat c^{\dagger}_{{\bf k}\lambda}(t)
\hat{c}_{{\bf k}\lambda}(t)
\;,\;
\hat H_{AI}
=
-
\int d{\bf r}\,
\hat {\bf P}({\bf r},t)
{\bf E}_{in}({\bf r},t)
\;,
\nonumber\\
\hat H_{AF}
&=&
-
\int d{\bf r}\,
\hat {\bf P}({\bf r},t)
\hat {\bf D}_{mic}({\bf r},t)
\;,
\nonumber
\end{eqnarray}
where the operator of the microscopic displacement field is given by
\begin{equation}
\label{dis-mic}
\hat {\bf D}_{mic}({\bf r},t)
=
\sum_{{\bf k}\lambda}
i
\sqrt{\frac{2\pi\hbar\omega_k}{V}}
{\bf e}_\lambda
\hat c_{{\bf k}\lambda}
\exp
\left(
      i {\bf k} {\bf r}
\right)
+ H.c.
\;,
\end{equation}
and the operator of the polarization field has the following form
\begin{equation}
\label{P}
\hat {\bf P}=
\sum_{j=1}^2
\hat {\bf P}_j
=
\sum_{j=1}^2
\left(
     \hat {\bf P}_j^+
     +
     \hat {\bf P}_j^-
\right)
=
\sum_{j=1}^2
{\bf d}_j
\left(
 \hat\psi_{gj}^\dagger \hat\psi_{ej}
 +
 \hat\psi_{ej}^\dagger \hat\psi_{gj}
\right)
\;.
\end{equation}
Here we assume that the incident electric field ${\bf E}_{in}({\bf r},t)$ is
produced by the laser, so it can be treated as a c-number function.
In the Hamiltonian (\ref{ham-second-quant}) we neglected all types of contact
interaction. This approximation is valid when the saturation parameters of
atomic transitions are small enough~\cite{WAL97,KBA99}. We do not include
into the Hamiltonian (\ref{ham-second-quant}) a trapping potential, because
our aim is to develop a theory of nonlinear atom optical processes of
unconfined atomic beams.

%--------------------------------------------------------------
\section{Heisenberg equations of motion for the atomic and
	 photonic operators}
%--------------------------------------------------------------

The Heisenberg equations of motion for the atomic and photonic operators
are easily derived by from the Hamiltonian (\ref{ham-second-quant}) and 
are given by:
\begin{eqnarray}
\label{heis-g}
&&
i \hbar
\frac{\partial\hat\psi_{gj}({\bf r},t)}{\partial t}
=
-
\frac{\hbar^2 \nabla^2}{2m_j}
\hat{\psi}_{gj}({\bf r},t)
-
{\bf d}_j
{\bf E}_{in}({\bf r},t)
\hat{\psi}_{ej}({\bf r},t)
\\
&&-
\hbar \sum_{{\bf k}\lambda} g_{{\bf k}\lambda j}^*
\hat{c}^{\dagger}_{{\bf k}\lambda}(t)
\exp
\left(
     - i {\bf k} {\bf r}
\right)
\hat{\psi}_{ej}({\bf r},t)
-
\hbar
\hat{\psi}_{ej}({\bf r},t)
\sum_{{\bf k}\lambda}
g_{{\bf k}\lambda j}
\exp\left( i {\bf k} {\bf r} \right)
\hat{c}_{{\bf k}\lambda}(t)
\;,
\nonumber
\end{eqnarray}
\begin{eqnarray}
\label{heis-e}
&&
i \hbar \frac{\partial\hat{\psi}_{ej}({\bf r},t)}{\partial t}
=
-
\frac{\hbar^2 \nabla^2}{2m_j}
\hat{\psi}_{ej}({\bf r},t)
+
\hbar \omega_j
\hat{\psi}_{ej}({\bf r},t)
- {\bf d}_j {\bf E}_{in}({\bf r},t)
\hat{\psi}_{gj}({\bf r},t)
\\
&&-
\hbar
\sum_{{\bf k}\lambda}
g_{{\bf k}\lambda j}^*
\hat{c}^{\dagger}_{{\bf k}\lambda}(t)
\exp
\left(
      - i {\bf k} {\bf r}
\right)
\hat{\psi}_{gj}({\bf r},t)
-
\hbar
\hat{\psi}_{gj}({\bf r},t)
\sum_{{\bf k}\lambda}
g_{{\bf k}\lambda j}
\exp
\left(
      i {\bf k} {\bf r}
\right)
\hat{c}_{{\bf k}\lambda}(t)
\;,
\nonumber
\end{eqnarray}
\begin{equation}
\label{heis-c}
i \hbar
\frac{\partial \hat c_{{\bf k}\lambda}(t)}{\partial t}
=
\hbar \omega_k
\hat c_{{\bf k}\lambda}(t)
-
\hbar 
\sum_{j=1}^2
g_{{\bf k}\lambda j}^*
\int d{\bf r}
e^{- i {\bf k} {\bf r}}
\left[
     \hat{\psi}_{ej}^\dagger({\bf r},t)
     \hat{\psi}_{gj}({\bf r},t)
     +
     \hat{\psi}_{gj}^\dagger({\bf r},t)
     \hat{\psi}_{ej}({\bf r},t)
\right]
\;,
\end{equation}
where ${\bf E}_{in}^\pm$ are the positive and negative frequency parts of the
incident classical electric field. The operator products in
Eqs.(\ref{heis-g}),(\ref{heis-e}),(\ref{heis-c}) are taken in normally
ordered form.

The formal solution of (\ref{heis-c}) for the photon operators is
\begin{eqnarray}
\label{solution-photon}
\hat{c}_{{\bf k}\lambda}(t)
=
\hat c_{{\bf k}\lambda}(0)
\exp
\left(
     - i \omega_k t
\right)
     &+& 
     i 
     \sum_{j=1}^2
     g_{{\bf k}\lambda j}^*
     \int_0^t
     dt'
     \int d{\bf r}'
     \exp
     \left[
	  i \omega_k (t'-t) - i {\bf k} {\bf r}'
     \right]
\\
&\times&     
     \left[
	  \hat \psi_{ej}^\dagger({\bf r}',t')
	  \hat{\psi}_{gj}({\bf r}',t')
	  +
	  \hat{\psi}_{gj}^\dagger({\bf r}',t')
	  \hat{\psi}_{ej}({\bf r}',t')
     \right]
\;,
\nonumber
\end{eqnarray}
where the first term $\hat{c}_{{\bf k}\lambda}(0)$ refers to the free-space
photon field and the second one goes back to the interaction with the atoms.

To study the back reaction of the photons on matter we insert
(\ref{solution-photon}) in (\ref{heis-g}) and (\ref{heis-e}).
By doing this procedure we eliminate photons in favor of atoms.
In the rotating-wave approximation we obtain the following dynamical equations
for the operators of the matter fields
\begin{eqnarray}
\label{matter-g}
i \hbar 
\frac{\partial\hat{\psi}_{gj}({\bf r},t)}{\partial t}
&=&
-
\frac{\hbar^2 \nabla^2}{2m_j} \hat{\psi}_{gj}({\bf r},t)
-
{\bf d} \hat{{\bf E}}_{loc}^-({\bf r},t)
\hat{\psi}_{ej}({\bf r},t)
\;,
\\
\label{matter-e}
i \hbar 
\frac{\partial\hat{\psi}_{ej}({\bf r},t)}{\partial t}
&=&
-
\frac{\hbar^2 \nabla^2}{2m_j} \hat{\psi}_{ej}({\bf r},t)
+
\hbar
\left(
     \omega_j + \delta_j - i \gamma_j/2
\right)
\hat\psi_{ej}({\bf r},t)
\nonumber\\
&&
-
\hat\psi_{gj}({\bf r},t)
{\bf d} \hat{\bf E}_{loc}^+({\bf r},t)
\;,
\end{eqnarray}
where $\delta_j$ and $\gamma_j$ are the Lamb shift and the spontaneous
emission rate of a single atom in free space, respectively.
We have introduced the operator of the local electric field
\begin{eqnarray}
\label{def-e-local}
\hat{{\bf E}}_{loc}^+({\bf r},t)
=
&&
{\bf E}_{in}^+({\bf r},t)
+
i
\sum_{{\bf k}\lambda}
\sqrt{\frac{2\pi\hbar\omega_k}{V}}
{\bf e}_\lambda
\hat{c}_{{\bf k}\lambda}(0)
\exp
\left(
     i {\bf k} {\bf r} - i \omega_k t
\right)
\nonumber\\
&&
+
\int
d{\bf r}'
\nabla \times \nabla \times
\frac { \hat{\bf P}^+ \left( {\bf r}',t-R/c \right)}{R}
\;,
\end{eqnarray}
where $\nabla\times$ refers to the point ${\bf r}$.
The polarization operator $\hat {\bf P}$ is given by eq.(\ref{P}).
Note that in Eq.(\ref{def-e-local}) a small volume around the
observation point ${\bf r}$ is excluded from the integration.

Eq. (\ref{def-e-local}) shows that $\hat{{\bf E}}_{loc}^\pm({\bf r},t)$ 
is a superposition of the incident field ${\bf E}_{in}^\pm({\bf r},t)$, 
vacuum fluctuations of the photon field, and the electric field
radiated by all other atoms, which has exactly the same form as in 
classical optics. It is this local field which
drives the inner atomic transition in Eqs.(\ref{matter-g}),
(\ref{matter-e})
which can be regarded as an atom-optical
analogue of the optical Bloch equations~\cite{ALL78,BOW93}.
They describe the dynamical evolution of second quantized matter
in the field of electromagnetic radiation.

%----------------------------------------------------------
\section{Lorentz-Lorenz relation and the system of
Maxwell-Bloch equations in atom optics of two-component BEC}
%----------------------------------------------------------

\subsection{Local-field correction}

The solution of Eqs. (\ref{matter-g}), (\ref{matter-e}) represents
a rather complicated mathematical problem because these equations
contain explicitly dipole-dipole interactions. In many particular situations
such a detailed microscopic description of matter is not necessary and it 
is more convenient to consider optical properties of the medium on a
macroscopic level. This can be done by introducing the macroscopic field
$\hat{{\bf E}}_{mac}({\bf r},t)$,
which 
satisfyes to
the macroscopic Maxwell equations for a charge-free and current-free
polarization medium,
instead of the local field
$\hat{{\bf E}}_{loc}({\bf r},t)$ in Eqs. (\ref{matter-g}), (\ref{matter-e}).

As in Ref.~\cite{GS} we can introduce the macroscopic field by setting
\begin{equation}
\label{local-corr}
\hat{{\bf E}}_{loc}^\pm({\bf r},t) =
\hat{{\bf E}}_{mac}^\pm({\bf r},t) +
\frac{4\pi}{3}\hat{{\bf P}}^\pm({\bf r},t)
\;.
\end{equation}
This equation is often called in the literature the Lorentz-Lorenz relation. 
It constitutes the basis of the
local-field effects in 
classical~\cite{BOR68}, 
quantum~\cite{GL}
and nonlinear optics (see \cite{BOW93,CB96,KM} and references therein).
In the case
of a classical electromagnetic field interacting with a macroscopic
dielectric medium this relation can be derived from first principles
under the assumption of homogeneity and isotropy of the dielectric
medium. We take it here as the definition of
$\hat{{\bf E}}_{mac}^\pm({\bf r},t)$.
It can then be shown with Eqs.(\ref{P}) and (\ref{def-e-local}) that this
$\hat{{\bf E}}_{mac}^\pm({\bf r},t)$ 
satisfies the macroscopic Maxwell equations, which can be written
down in the form of the wave equation
\begin{equation}
\label{wave-equation}
\nabla \times \nabla \times 
\hat{{\bf E}}_{mac}^\pm({\bf r},t) 
= 
- \frac{1}{c^2} 
\frac{\partial^2 \hat{{\bf E}}_{mac}^\pm({\bf r},t)}{\partial t^2} 
-
\frac{4\pi}{c^2} 
\frac{\partial^2 \hat{{\bf P}}^\pm({\bf r},t)}{\partial t^2}
\;,
\end{equation}
so it is justified to call it the quantum field operator of 
the macroscopic electric field. At the same time
this definition allows us 
to interpret our results on ultracold atomic gases in analogy to
the interaction of light with a macroscopic dielectric medium. 

%---------------------------------------------------------------------
\subsection{Nonlinear matter equation}
%---------------------------------------------------------------------

We substitute (\ref{local-corr}) in (\ref{matter-g}) and
(\ref{matter-e}) and pass to a reference frame rotating with frequency
$\omega_L$ of the incident field, which is assumed to be monochromatic,
to obtain
\begin{equation}
\label{g1}
i \hbar
\frac{\partial \hat \psi_{g1}}{\partial t}
=
-
\frac{\hbar^2 \nabla^2}{2m_1}
\hat \psi_{g1}
-
\frac{\hbar}{2}
\hat{\Omega}_1^-({\bf r})
\hat \phi_{e1}
-
\frac{4\pi}{3}
d_1^2
\hat \phi_{e1}^\dagger
\hat \psi_{g1}
\hat \phi_{e1}
-
\frac{4\pi}{3}
{\bf d}_1 {\bf d}_2
\hat \phi_{e2}^\dagger
\hat \psi_{g2}
\hat \phi_{e1}
\;,
\end{equation}
\begin{eqnarray}
\label{e1}
i \hbar
\frac{\partial\hat{\phi}_{e1}}{\partial t}
&=&
-
\frac{\hbar^2 \nabla^2}{2m_1}
\hat{\phi}_{e1}
-
\frac{\hbar}{2}
\hat{\psi}_{g1}
\hat{\Omega}_1^+({\bf r})
- \frac{4\pi}{3} d_1^2
\hat{\psi}_{g1}
\hat{\psi}_{g1}^\dagger
\hat{\phi}_{e1}
\nonumber
\\
&&
-
\hbar
\left(
\Delta_1 + i \gamma_1/2
\right)
\hat{\phi}_{e1}
-
\frac{4\pi}{3}
{\bf d}_1 {\bf d}_2
\hat \psi_{g1}
\hat \psi_{g2}^\dagger
\hat \phi_{e2}
\;,
\end{eqnarray}
\begin{equation}
\label{g2}
i \hbar
\frac{\partial \hat \psi_{g2}}{\partial t}
=
-
\frac{\hbar^2 \nabla^2}{2m_2}
\hat \psi_{g2}
-
\frac{\hbar}{2}
\hat{\Omega}_2^-({\bf r})
\hat \phi_{e2}
-
\frac{4\pi}{3}
d_2^2
\hat \phi_{e2}^\dagger
\hat \psi_{g2}
\hat \phi_{e2}
-
\frac{4\pi}{3}
{\bf d}_1 {\bf d}_2
\hat \phi_{e1}^\dagger
\hat \psi_{g1}
\hat \phi_{e2}
\;,
\end{equation}
\begin{eqnarray}
\label{e2}
i \hbar
\frac{\partial\hat{\phi}_{e2}}{\partial t}
&=&
-
\frac{\hbar^2 \nabla^2}{2m_2}
\hat{\phi}_{e2}
-
\frac{\hbar}{2}
\hat{\psi}_{g2}
\hat{\Omega}_2^+({\bf r})
- \frac{4\pi}{3} d_2^2
\hat{\psi}_{g2}
\hat{\psi}_{g2}^\dagger
\hat{\phi}_{e2}
\nonumber\\
&&
-
\hbar
\left(
\Delta_2 + i \gamma_2/2
\right)
\hat{\phi}_{e2}
-
\frac{4\pi}{3}
{\bf d}_1 {\bf d}_2
\hat \psi_{g2}
\hat \psi_{g1}^\dagger
\hat \phi_{e1}
\;,
\end{eqnarray}
with the detunings $\Delta_j=\omega_L-\omega_j-\delta_j$, $j=1,2$. 
The position dependent Rabi frequencies
$
\hat{\Omega}^\pm_j({\bf r})
=
2{\bf d}_j\hat{\bf \cal E}_{mac}^\pm({\bf r})/\hbar
$
are related to the macroscopic electric field.

Because we are mainly interested in atom optical problems and want
to study the coherent evolution of the center-of-mass motion of the gas,
we shall neglect spontaneous emission. This is valid for situations
where the absolute values of the detunings are much bigger than
the spontaneous emission rates and Rabi frequencies 
$
\left|\Delta_j\right| \gg \gamma_j
\;,\;
\left|
     \Omega_j
\right|
$.
In order to do this approximation self-consistently we drop in 
the following the vacuum fluctuations
and the spontaneous emission rates $\gamma_j$ from our equations.
In addition we shall replace all the operators by macroscopic
functions.
We may, therefore, apply
{\it the adiabatic approximation}~\cite{ZHA94a,ALL78,MZ98} 
to (\ref{e1}), (\ref{e2}), which gives
\begin{equation}
\label{adiabatic-sol}
{\phi}_{ej}({\bf r},t) =
-
\frac
{{\Omega}_j^+({\bf r}) {\psi}_{gj}({\bf r},t)}
{ 
  2 \Delta_j^{loc}({\bf r},t) 
}
\;,
\end{equation}
where the local detuning is given by
\begin{equation}
\label{loc-det}
\Delta_j^{loc}({\bf r},t)
=
\Delta_j
\left\{
  1
  -
  \frac{4\pi}{3}
  \left[
     \alpha_1
     \left|
	 \psi_{g1}({\bf r},t)
     \right|^2
     +
     \alpha_2
     \left|
	 \psi_{g2}({\bf r},t)
     \right|^2
  \right]
\right\}
\;.
\end{equation}
Here
$\alpha_j=-d_j^2/\hbar\Delta_j$ is the atomic polarizability for
$j$-th component.

Then substituting (\ref{adiabatic-sol}) in (\ref{g1}), (\ref{g2}),
which eliminates the excited states, we obtain as the
result a system of nonlinear
equations for the ground state matter fields ${\psi}_{gj}({\bf r},t)$
\begin{equation}
\label{nonlinear-equation}
i\hbar
\frac{\partial \psi_{gj}({\bf r},t)}{\partial t}
=
\left\{
   -
   \frac{\hbar^2\nabla^2}{2m_j}
   +
   \frac
   {\hbar \Delta_j \left|  \Omega_j^+({\bf r})\right|^2}
   {
     4
     \left[
         \Delta_j^{loc}({\bf r},t)
     \right]^2
   }
\right\}
\psi_{gj}({\bf r},t)
\;.
\end{equation}

Varying the parameters in Eq.(\ref{nonlinear-equation}) we can change
the nonlinear potential which is given by the second term on the r.h.s.
of Eq.(\ref{nonlinear-equation}).
For instance, for increasing densities and positive detunings
$\Delta_j$ the local detunings grow and, correspondingly, the nonlinear term
in (\ref{nonlinear-equation}) representing the coupling to the macroscopic
electric field becomes smaller. On the other hand, for negative detunings
the absolute values of the local detunings decrease with increasing
densities and the nonlinearity becomes greater. This behavior is
exactly the same as we had in a one-component medium. In a two-component
medium another regime is possible which can not be reached in a one-component
medium: If the signs of the detunings are different, then one can increase
the densities of the components in such a manner that the values of the local
detunings, and therefore of the nonlinear potentials, will remain constant.

While Eq.(\ref{nonlinear-equation}) will allow us to derive an expression 
for the
dielectric susceptibility of a Bose gas which closely resembles that
of a classical gas we have to remark that it is only valid for low enough
values of parameters 
$
  \varepsilon_j
  =
  \alpha_j
  \left|
       \psi_{gj}
  \right|^2
$. 
The reason is that the adiabatic approximation 
(\ref{adiabatic-sol})
represents the first-order term in an expansion in $1/\Delta_j$~\cite{ALL78}. 
Therefore one also should expand Eq.~(\ref{nonlinear-equation}) to
first order in this parameter. This procedure leads to a pair of
coupled Gross-Pitaevskii equations
\begin{equation}
\label{GP}
i\hbar
\frac{\partial  \psi_{gj}}{\partial t}
=
\left\{
   -
   \frac{\hbar^2\nabla^2}{2m_j}
   +
   \frac{\hbar}{4\Delta_j}
   { \left|  \Omega_j^+\right|^2 }
   {
     \left[
       1
       +
       \frac{8\pi}{3}
       \left(
	  \alpha_1
	  \left|
	       \psi_{g1}
	  \right|^2
	  +
	  \alpha_2
	  \left|
	       \psi_{g2}
	  \right|^2
       \right)
     \right]
   }
\right\}
 \psi_{gj}
\;.
\end{equation}
Equations of this type have been used, for instance, 
in papers~\cite{PHTS,BGH}.

%-----------------------------------------------------------------
\subsection{\bf Optical properties of the two-component ultracold gas}
%-----------------------------------------------------------------

Making use of the adiabatic solutions (\ref{adiabatic-sol}) we
obtain the following expression for the medium polarization
\begin{equation}
\label{media-equ}
{{\bf P}}^+({\bf r},t)
=
{\chi}({\bf r},t) {{\bf E}}_{mac}^+
({\bf r},t)
\;,
\end{equation}
where dielectric susceptibility is given by
\begin{equation}
\label{def-suscept}
{\chi}({\bf r},t)=
\frac
{
  \sum_{j=1}^2
  \alpha_j
  \left|
       \psi_{gj} ({\bf r},t)
  \right|^2
}
{
  1
  -
  \frac{4\pi}{3}
  \sum_{j=1}^2
  \alpha_j
  \left|
       \psi_{gj} ({\bf r},t)
  \right|^2
}
\;.
\end{equation}
Dielectric susceptibility is a rather important parameter, because it
describes the propagation of the laser radiation inside a medium. In most
of the practical situations the electromagnetic processes are much faster
than the center-of-mass motion of the atoms. Therefore, $\chi$ can be
considered as a time-independent quantity.
Let us assume in addition that the spatial variations of the atomic density 
are not very large, such that $\nabla \chi \to 0$. Then 
$\mbox{div} \, {\bf E}_{mac}^\pm \approx 0$, and we have the following Helmholtz
equation for the macroscopic electric field
\begin{equation}
\label{maxwell-media}
\nabla^2 {\bf \cal E}_{mac}^\pm  +
k_L^2 {n}^2 {\bf \cal E}_{mac}^\pm = 0
\;,
\end{equation}
with the refractive index $n$ given by the Maxwell-Garnett formula
\begin{equation}
\label{MG}
{n}^2 
=
1 + 4 \pi \chi
=
\frac
{
  1 
  + 
  \frac{8\pi}{3} 
  \sum_{j=1}^2 
  \alpha_j
  \left|
       \psi_{gj}
  \right|^2
}
{
  1 
  - 
  \frac{4\pi}{3} 
  \sum_{j=1}^2 
  \alpha_j
  \left|
       \psi_{gj}
  \right|^2
}
\;,
\end{equation}
which is a two-component analogue of the Clausius-Mossotti formula.

Eqs. (\ref{nonlinear-equation}), (\ref{maxwell-media}), (\ref{MG})
can be considered as an atom optical analogue of the system of
Maxwell-Bloch equations.
In general they have to be solved in a self-consistent 
way and in usual situations solutions can be obtained only by doing
numerical calculations. In the next section we shall consider
one particular example of an analytical description of a nonlinear
atom optical system.

%-----------------------------------------------------------------------
\section
{
  Diffraction of a two-component ultracold atomic beam from a strong 
  standing light wave
}
%-----------------------------------------------------------------------

We consider a typical scheme for the observation of diffraction in atom optics: 
An incident atomic beam moves in 
$z$-direction, perpendicular to two laser waves
counter propagating along the $y$-axis with wave vectors $+{\bf k}_L$
and $-{\bf k}_L$, respectively, and with Gaussian envelope.
From the uncertainty relation it follows that in order to
get a distinct diffraction pattern, the width of the atomic 
wave packet $w_y$ should be sufficiently large compared to the wave lengh of the laser 
radiation in a medium. In this case the atoms can be described as a 
homogeneous medium with constant refractive index. 
If the spontaneous emission does not make any contribution,
the effect of the atoms on the laser beam is purely 
dispersive and only the wavelength will be shifted. 
This means that in a medium we shall have a standing wave which is formed
by counter propagating laser beams with the wave vectors $+n{\bf k}_L$
and $-n{\bf k}_L$, respectively.
In this approximation the solution of 
(\ref{maxwell-media}) with (\ref{MG}) is given by
\begin{equation}
\label{sol-maxwell}
\left| \Omega^+_j \right|^2 
=
\left| \Omega_j \right|^2
\exp
\left( 
     - z^2 / w_L^2 
\right)
\cos^2 nk_L y
\;. 
\end{equation}
We assume that the longitudinal kinetic energy of the atomic beam,
associated with the center-of-mass motion in $z$ direction, is large
compared to the nonlinear potential in eq.(\ref{nonlinear-equation}).
Then the $z$-component of the atomic velocity will not change much
and, therefore, the motion of atoms in $z$ direction during the whole
evolution can be treated classically. Only the motion in $y$ direction
should be treated quantum mechanically. In such a situation the coordinate
$z$ plays the role of time and we can change the variable $t=z/v_{gj}$ in
(\ref{nonlinear-equation}) with $v_{gj}$ being the group velocity of the
$j$-th component.
In addition we assume that we are in the Raman-Nath regime and we can
neglect the transverse kinetic energy during the
interaction of the atoms with the electromagnetic field. This approximation
is valid for heavy atoms or if the interaction is so strong that atoms can
take up momentum without changing considerably the velocity~\cite{Sch}. In this case
the density of atoms remains unaltered, but their phase changes.
Making use of all these assumptions we can write down
the solutions of eqs.(\ref{nonlinear-equation}) for $z \gg w_L$
(in the far zone) in the following form
\begin{equation}
\label{sol-diffract-1}
{\psi}_{gj}(y,\infty)
=
{\psi}_{gj}(y,-\infty)
\exp
\left(
\int_{-\infty}^\infty
\frac
{-i\left| \Omega_j^+(y,z) \right|^2}
{4\Delta_j v_{gj}\left( 1 + V_1 \rho_{g1} + V_2 \rho_{g2}\right)^2} 
dz 
\right)
\;,
\end{equation}
where 
\begin{equation}
\label{eff-volume}
V_j 
=
-
\frac{4\pi}{3}
\alpha_j
=
\frac{4\pi}{3\hbar} 
\frac{d_j^2}{\Delta_j} 
\;, 
\end{equation}
and
$
\rho_{gj}
=
\left|
     \psi_{gj}
\right|^2
$
is the density of
atoms in the ground state.

We represent $\rho_{gj}$ as Gaussian wave packets with width $w_y$
\begin{equation}
\label{density-profile}
\rho_{gj} 
= 
\rho_j
\exp
\left( 
    - y^2 / w_y^2 
\right) 
\;.
\end{equation}
Then we substitute Eqs.(\ref{sol-maxwell}) and (\ref{density-profile}) into
Eq.(\ref{sol-diffract-1}) and take into account that the width of the atomic
wave packet must be much larger than the
wavelenght of the laser radiation, i.e., $w_y \gg 2\pi/ nk_L$.
After integration we get the following result
\begin{equation}
\label{expansion}
{\psi}_{gj}(y,\infty) =
{\psi}_{gj} \left(y,-\infty\right)
e^{- i \tau_j}
\sum_{q=-\infty}^\infty
e^{i 2 q nk_L y } (-i)^q J_q(\tau_j)
\;,
\end{equation}
which is represented here in the form of a Fourier series expansion. We use
the notations:
\begin{equation}
\tau_j = 2 g_j / \left( 1 + V_1 \rho_1 + V_2 \rho_2 \right)^2
\;,\;
g_j = \frac{\Omega_j^2}{16\Delta_j} \frac{w_L}{v_{gj}} \sqrt{\pi}
\;.
\end{equation}
$J_q$ is the $q$-th order Bessel function.

From the solution (\ref{expansion}) it follows that the momentum transferred
from the laser beam to the atomic beam is the same for both components
and equals to $2 q nk_L y$. It is determined by the wave number of the
incident laser radiation $k_L$ and the refractive index of the gas $n$.
However,
the probabilities to find the components of the beam in a momentum state
$2 q n k_L$ are different for different components:
\begin{equation}
\label{probabilities}
P_{qj} = J_q^2(\tau_j)
\;,\quad
q=0, \pm 1, \pm 2, ...,
\end{equation}
with $P_{0j}$ being the probability to find the $j$-th component
in the same momentum state as for the incident atomic beam. The angle of
diffraction $\alpha_{qj}$ for a particular momentum state $q$ and
for a particular component $j$ is thereby given by
\begin{equation}
\label{angles}
\tan \alpha_{qj} = \frac{2 q n \hbar k_L}{m_j v_{gj}}
\;.
\end{equation}

Therefore the diffraction pattern, as it follows from Eqs.(\ref{expansion}),
(\ref{probabilities}), (\ref{angles}),
depends on the densities of the components.
Depending on the values of $m_j$ and $v_{gj}$ the angle $\alpha_{1q}$
can be either the same as $\alpha_{q2}$ or different. Only if
$m_1 v_{g1} = m_2 v_{g2}$, i.e., when the momenta of different components
associated with the group velocities are the same,
$\alpha_{q1} = \alpha_{q2}$. In all other situations, for instance,
if the group velocities of the components are equal to each other or if
we have a monoenergetic atomic beam, $\alpha_{q1} \ne \alpha_{q2}$,
and in the diffraction pattern one can observe spatially separated
components.

%-----------------------------------------------------------------
\section{Conclusion}
%-----------------------------------------------------------------

In the present paper we have investigated the process of the interaction
of a two-component BEC with the field of vacuum and laser photons.
The two-component BEC is treated as a binary mixture of two-level atoms
with different masses, transition frequencies and transition dipole
moments. Starting from the microscopic model and making use of the
multipolar formulation of QED, a general system of Maxwell-Bloch
equations is derived which can be used for the description of
nonlinear phenomena in atom optics. Optical properties of the
two-component BEC are investigated. The refractive index is shown to satisfy
the Maxwell-Garnett formula. 

As a typical atom optical application,
we have considered the diffraction of two-component atomic beam from
a strong standing laser wave in the Raman-Nath approximation, which
allows to obtain simple analytical solutions. It is shown that in
most of the situations one can observe splitted components of the beam
in the diffraction pattern. 

The limits of validity of the results, obtained in the present paper,
are essentially restricted by the adiabatic approximation, which is
correct up to the first order with respect to the small parameters 
$1/\Delta_j$. Therefore, our results are valid for small 
enough
$
  \varepsilon_j
  =
  \alpha_j
  \left|
       \psi_{gj}
  \right|^2
$
.
They generalize our previous results~\cite{KBA99,KBA00}. 

Although, we have considered
here explicitly only a two-component BEC, the generalization
to an arbitrary number of different atomic species is
straightforward and can be done very easily.

%----------------------------------------------------------------
\section*{Acknowledgments}
%----------------------------------------------------------------

This work has been supported by the Deutsche For\-schungsgemeinschaft
and the Optikzentrum Konstanz. One of us (K.V.K.) would like to
thank also the Alexander-von-Humboldt Stiftung for financial support.
This work has been partly inspired by the discussions with C.M.Bowden
and M.Crenshaw.

\newpage

%---------------------------------------------------------------------

\end{document}